\documentclass[nologo,10pt,a4paper]{ETHpaper}
\usepackage{graphicx,amsmath,amssymb}\usepackage{dcolumn}\usepackage{bm}\usepackage{subfigure}
\usepackage{cleveref}

\DeclareRobustCommand{\mean}[1]{\left\langle #1 \right\rangle}
\newcommand{\abs}[1]{\left| #1 \right|}

\begin{document}
\title{From Relational Data to Graphs: \\Inferring Significant Links using\\ Generalized Hypergeometric Ensembles}
\titlealternative{From Relational Data to Graphs: Inferring Significant Links}

\author{Giona Casiraghi$^{*}$, Vahan Nanumyan, Ingo Scholtes, Frank Schweitzer}
\authoralternative{G. Casiraghi, V. Nanumyan, I. Scholtes, F. Schweitzer}
\address{$^*$Corresponding author; E-mail: gcasiraghi@ethz.ch\\
Chair of Systems Design, ETH Zurich, 
Weinbergstrasse 56/58, CH-8092 Zurich, Switzerland}
\www{\url{http://www.sg.ethz.ch}}
\reference{}
\date{\today}
\makeframing
\maketitle

\begin{abstract}
\noindent The inference of network topologies from relational data is an important problem in data analysis.
Exemplary applications include the reconstruction of social ties from data on human interactions, the inference of gene co-expression networks from DNA microarray data, or the learning of semantic relationships based on co-occurrences of words in documents.
Solving these problems requires techniques to infer significant links in noisy relational data.
In this short paper, we propose a new statistical modeling framework to address this challenge.
It builds on \emph{generalized hypergeometric ensembles}, a class of generative stochastic models that give rise to analytically tractable probability spaces of directed, multi-edge graphs.
We show how this framework can be used to assess the significance of links in noisy relational data.
We illustrate our method in two data sets capturing spatio-temporal proximity relations between actors in a social system.
The results show that our analytical framework provides a new approach to infer significant links from relational data, with interesting perspectives for the mining of data on social systems.

\end{abstract}

\section{Motivation}

Advances in data sensing and collection give rise to an increasing volume of data that capture \emph{dyadic relations} between elements or actors in social, natural, and technical systems.
While it is common to apply \emph{graph mining} and \emph{network analysis} to such relational data, it is often questionable whether the application of these techniques is actually justified.
Consider, for instance, various forms of \emph{time series data}, which not only tell us which elements of a complex system are related but also when or in which order relations occur.
Such data give rise to \emph{temporal networks}, which question the application of widely used network-based modeling and data mining techniques~\cite{Rosvall2014,Holme2015,Scholtes2016,Vidmer2016,Scholtes2017}.
Apart from temporal information, we often have access to data that capture multiple types of relations or interactions.
The resulting \emph{multi-layer network topologies} give rise to complications that threaten standard techniques, e.g., to infer and analyze social networks, detect community structures, or to model and control dynamical processes in networked systems~\cite{szell2010,Kivela2014,Domenico2015,ZhangY2016,Casiraghi2017}.

The challenges outlined above are due to the growing availability of \emph{additional information} -- such as time-stamped, sequential or multi-dimensional relational data -- which must be incorporated into network-based techniques to model and analyze relational data.
However, we are often confronted with situations in which we \emph{lack information} that is needed to interpret observed relations.
Consider, for instance, data sets that capture the simultaneous presence of two users at the same location, the joint expression of two genes in a DNA microarray, or the co-occurrence of two words in the same document.
Each of these observed relations can either be due to an underlying social tie, a functional relationship between genes, a semantic link between two words, or it could simply have occurred by mere chance.
Rather than na\^ively analyzing such data from the perspective of graphs or networks, we should thus treat them as \emph{noisy observations} that may or may not indicate true \emph{relations} between a system's elements.

Principled and efficient methods to solve this \emph{network inference problem} are of major importance for the modeling and analysis of social networks, the reconstruction of biological networks, and the mining of semantic structures in information systems.
The problem has received significant attention from the data mining and machine learning community, as well as from researchers in graph theory and network science.
Especially in the latter community, the problem is commonly addressed using \emph{statistical ensembles}, i.e., generative stochastic models of graphs that can be used for inference, learning and modeling tasks.
A common issue of these techniques is that the underlying statistical ensembles are not analytically tractable, thus requiring time-consuming numerical simulations and Monte-Carlo sampling techniques.

To address this problem, in this short paper we propose \emph{generalized hypergeometric ensembles} (gHypE), a novel framework of statistical ensembles to infer significant links in relational data.
The framework can be viewed as generalization of the configuration model, which is commonly used to generate random graph topologies with a given sequence of node degrees.
Our framework extends this state-of-the-art graph-theoretic approach in two ways.
First, it provides \emph{analytically tractable} probability spaces of directed and undirected multi-edge graphs, eliminating the need for expensive numerical simulations.
Second, it allows to account for known factors that influence the occurrence of interactions, such as known group structures, similarities between elements, or other forms of biases.
We demonstrate our framework in two real-world data sets that capture spatio-temporal proximities of actors in a social system.
The results show that our framework provides interesting new perspectives for the mining and learning in graphs.

\section{Background and Related Work}

The problem of inferring \emph{significant} links in relational data has been addressed in a number of works.
In the following, we coarsely categorize them into three lines of research.

Applying predictive analytics techniques, a first set of works studied the problem from the perspective of \emph{link prediction}~\cite{Liben-Nowell2007}.
In \cite{Tang2012}, a supervised learning technique is used to predict \emph{types} of social ties based on unlabeled interactions.
The authors of \cite{Schein2015} show that tensor factorization techniques allow to infer international relations from data that capture how often two countries co-occur in news reports.
In \cite{Xiang2010}, a link-based latent variable model is used to predict friendship relations using data on social interactions.

Using the special characteristics of time-stamped social interactions or geographical co-occurrences, a second line of works has additionally accounted for \emph{spatio-temporal information}.
Studying data on time-stamped proximities of students at MIT campus, the authors of \cite{Eagle2009} show that the temporal and spatial distribution of proximity events allows to infer social ties with high accuracy.
In \cite{Cranshaw2010}, a model that captures location diversity, regularity, intensity and duration is used to predict social ties based on co-location events.
An entropy-based approach taking into account the diversity of interactions' locations has been used in \cite{Pham2013}.

Addressing scenarios where neither training data nor spatio-temporal information is available, a third line of works is based on \emph{generative models for random graphs}.
Such models can be used as \emph{null models} for observed dyadic interactions, which help us to assess whether the relations between a given pair of elements occur significantly more often than expected.
Existing works in this area typically rely on standard modeling frameworks, such as \emph{exponential random graphs}~\cite{Robins2007,Cimini2014}, or the \emph{configuration model} for graphs with given degree sequence or distribution~\cite{Molloy1995}.
On the one hand, these approaches provide statistically principled network inference and learning methods for general relational data~\cite{Anand2009,Newman2015,Wilson2014,Gemmetto2017}.
On the other hand, the underlying generative models are often not analytically tractable, thus requiring expensive numerical simulations~\cite{Robins2007,Newman2015}.
Proposing a framework of analytically tractable generative models for directed and undirected multi-edge graphs, in this work we close this research gap.

\section{Generalized Hypergeometric Ensembles}

In the following we introduce our framework step by step.
For this, let us first consider a data set consisting of repeated dyadic interactions $(i,j)$, which have been observed between two nodes $i$ and $j$.
Such a data set can be represented as a \emph{multi-edge}, or \emph{weighted}, network $G=(V,E)$, where $V$ is a set of $n$ nodes, and $E \subseteq V \times V$ is a multi-set of (directed or undirected) edges.
Let us further define an adjacency matrix $\hat{\mathbf{A}}$, where entries $\hat{A}_{ij}\in\mathbb N_{0}$ capture the \emph{weight} of an edge $(i,j)\in V \times V$, i.e., the multiplicity of an edge $(i,j)$ in the multi-set $E$.
For each node $i \in V$ we further define the (weighted) in-degree $\hat{k}_{\mathrm{in}}(i) := \sum_{j \in V} \hat{A}_{ji}$ and the (weighted) out-degree $\hat{k}_{\mathrm{out}}(i) := \sum_{j \in V} \hat{A}_{ij}$.

Rather than directly applying graph mining and learning techniques to such a weighted graph $G$, in the following we are interested in a crucial question:
Which of the links between nodes are \emph{significant}, i.e., which of the observed weights $A_{ij}$ go beyond what is expected at random, given (i) the total number of observed interactions, and (ii) the number of times individual nodes engage in interactions?
To answer this question, we take the common approach of defining a \emph{stochastic model} that generates a so-called \emph{statistical ensemble}, i.e., a probability space of graphs.
Different from existing approaches, where link weights are assumed to be continuous (e.g. \cite{Aicher2015,DeChoudhury2010}), we are interested in a statistical ensemble that (i) can handle directed and multi-edge graphs, (ii) is analytically tractable, and (iii) thus allows us to assess the significance of links in a theoretically principled way.

Our construction of a statistical ensemble follows the general idea of the Molloy-Reed configuration model, which is to randomly shuffle the topology of a given network $G$ while preserving the observed node degrees.
For this, the configuration model generates edges between randomly sampled pairs of nodes in such a way that the \emph{exact} observed degrees of all nodes are preserved.
Different from this approach, we assume a sampling of $m$ multi-edges such that the sequence of \emph{expected} degrees of nodes is preserved.
For this, for each pair of nodes $i$ and $j$, we first define the maximum number $\Xi_{ij}$ of multi-edges that can possibly exist between nodes $i$ and $j$ as $\Xi_{ij} := \hat{k}_{\mathrm{out}}(i) \hat{k}_{\mathrm{in}}(j)$ (cf. \cite{Newman2006,Karrer2011}).
The maximally possible numbers of links between all pairs of nodes can then be conveniently represented in matrix form as $\mathbf{\Xi} := \left(\Xi_{ij}\right)_{i,j \in  V}$.

Our statistical ensemble is then defined by the following sampling procedure:
For each pair of nodes $i,j$, we sample edges from a set of $\Xi_{ij}$ possible multi-edges uniformly at random.
This can be viewed as an \emph{urn problem}~\cite{jacod2003probability} where the edges to be sampled are represented by balls in an urn.
By representing edges connecting different pairs of nodes $(i,j)$ as balls having $n^{2}=\abs{V\times V}$ different colours, we obtain an urn with a total of $M=\sum_{i,j}\Xi_{ij}$ differently colored balls.
With this, the sampling of a network according to our model corresponds to drawing exactly $m$ balls from this urn.
Each adjacency matrix $\mathbf{A}$, with entries $A_{ij}$ such that $\sum_{i,j} A_{ij}=m$, corresponds to one particular realization drawn from this ensemble.
The probability to draw exactly $\mathbf{A}=\{A_{ij}\}_{i,j \in  V}$ edges between each pair of nodes is given by the \emph{multivariate} hypergeometric distribution~\footnote{Note that we do not distinguish between  the $n\times n$ adjacency matrix $\mathbf{A}$ and the $n^{2}\times 1$ vector obtained by stacking.}
\begin{equation}
	\label{eq:hypergeometricNet}
  \Pr(\mathbf{A}) = \dbinom{M}{m}^{-1}\prod_{i,j}\dbinom{\Xi_{ij}}{A_{ij}}.
\end{equation}

For each pair of nodes $i,j \in V$, the probability to draw exactly $\hat{A}_{ij}$ edges between $i$ and $j$ is given by the marginal distributions of the multivariate hypergeometric distribution.
We thus arrive at a \emph{hypergeometric statistical ensemble}, which (i) generalizes the configuration model to directed, multi-edge graphs, (ii) has a fixed sequence of \emph{expected} degrees, and (iii) is analytically tractable.
Moreover, it provides a framework to generalize other random graph models like, e.g., the multi-edge version of the Erd\"os-R\'enyi model~\cite{erdds1959random}, where only $n$ and $m$ are fixed, while there are no constraints on the degree sequence.
This corresponds to a definition of $\mathbf{\Xi}$ with $\Xi_{ij}=m^{2}/n^{2}=\;$const. which directly results from $\mean{k_{\mathrm{in}}(i)}=\mean{k_{\mathrm{out}}(i)}=m/n$.

The sampling procedure above gives a stochastic model for weighted, directed graph in which (i) the expected weighted in- and out-degree sequence is fixed, and (ii) interactions between nodes are generated at random.
This provides a null model in which the probability for a particular pair of nodes to be connected by an edge is only influenced by combinatorial effects, and thus only depends on the node degrees.
For scenarios where we have additional information on factors that influence the formation of edges, we can further generalize the ensemble above as follows:
We introduce a matrix $\mathbf{\Omega}$ whose entries $\Omega_{ij}$ capture relative \emph{dyadic propensities}, i.e., the tendency of a node $i$ to form an edge \emph{specifically} to node $j$.
These propensities $\Omega_{ij}$ \emph{bias} the edge sampling process described above.
This implies that entry $\Omega_{ij}$ only captures the propensity that goes \emph{beyond} the tendency of a node $i$ to connect to a node $j$ that is due to combinatorial effects, i.e., the in-degree of $j$ and the out-degree of $i$.
In analogy to the urn model, here a biased sampling implies that the probability of drawing balls of a given color (representing all possible edges between a given pair of nodes) does not only depend on their number but also on the respective relative propensities.
The probability distribution resulting from such a biased sampling process is given by the multivariate \emph{Wallenius' non-central hypergeometric distribution}~\cite{Wallenius1963,Fog2008a}:
\begin{equation}
	\label{eq:walleniusNet}
	\Pr(\mathbf{A})=\left[\prod_{i,j}{\dbinom{\Xi_{ij}}{A_{ij}}}\right]
         \int_{0}^{1}{\prod_{i,j}{\left(1-z^{\frac{\Omega_{ij}}{S_{\mathbf{\Omega}} }}\right)^{A_{ij}}}dz}
\end{equation}
with
$S_{\mathbf{\Omega}}= \sum_{i,j} \Omega_{ij}(\Xi_{ij}-A_{ij})$.

Similar to the unbiased sampling described above,the probability to observe a particular number $\hat{A}_{ij}$ of edges between a pair of nodes $i$ and $j$ can again be calculated from the marginal distribution as
\begin{equation}
	\label{eq:walleniusEdge}
  \begin{aligned}
	\Pr(&A_{ij}=\hat{A}_{ij}) =  \dbinom{\Xi_{ij}}{\hat{A}_{ij}}\dbinom{M-\Xi_{ij}}{m-\hat{A}_{ij}}\cdot  \int_{0}^{1} \left[ \vphantom{e^\frac12}\right.\\
          &\left(
                    1 - z^{ \frac{\Omega_{ij}}{S_{\mathbf{\Omega}}}}
\right)^{\hat{A}_{ij}} \left(
                    1-z^{ \frac{\bar{\Omega}_{\setminus(i,j)}}{S_{\mathbf{\Omega}}}}
\right)^{m-\hat{A}_{ij}}
\left.\vphantom{\vphantom{e^\frac12}}\right]dz
  \end{aligned}
\end{equation}	
where
$\bar{\Omega}_{\setminus(i,j)} = (M-\Xi_{ij})^{-1}\sum_{(l,m)\in
  V\times V\backslash(i,j)}{\Xi_{lm}\Omega_{lm}}$.

Note that for the special case of a uniform dyadic propensity matrix $\mathbf{\Omega} \equiv \text{const}$, we recover Eq.~\ref{eq:hypergeometricNet} for the unbiased case, i.e., where all dyadic propensities are identical.
We thus obtain a general framework of statistical ensembles which (i) allows to encode arbitrary a priori tendencies of nodes to interact, and (ii) provides an analytical expression for the probability to observe a given number of interactions between any pair of nodes.

\section{Inferring Significant Social Ties}

In the following, we demonstrate how our framework can be used to infer significant links in two relational data sets:
(RM) captures time-stamped proximities between students and faculty at MIT~\cite{Eagle2006} recorded via smart devices.
(ZKC) covers frequencies of self-reported encounters between members of a university Karate club collected by Wayne Zachary~\cite{Zachary}.
We denote the weighted adjacency matrix capturing observed dyadic interactions as $\mathbf{\hat{A}}$.
For a given significance threshold $\alpha$, we then identify significant links by filtering matrix $\mathbf{\hat{A}}$ by a threshold $\Pr(A_{ij} \leq \hat{A}_{ij}) > 1 - \alpha$ based on Eq.~\ref{eq:walleniusEdge}.
This can be seen as assigning $p$-values to dyads $(i,j)$, obtaining a \emph{high-pass} noise filter for entries in the adjacency matrix.

\begin{figure}[htpb]
  \centering
  \subfigure[\label{fig:rm:unfiltered}]{\includegraphics[width=.28\columnwidth]{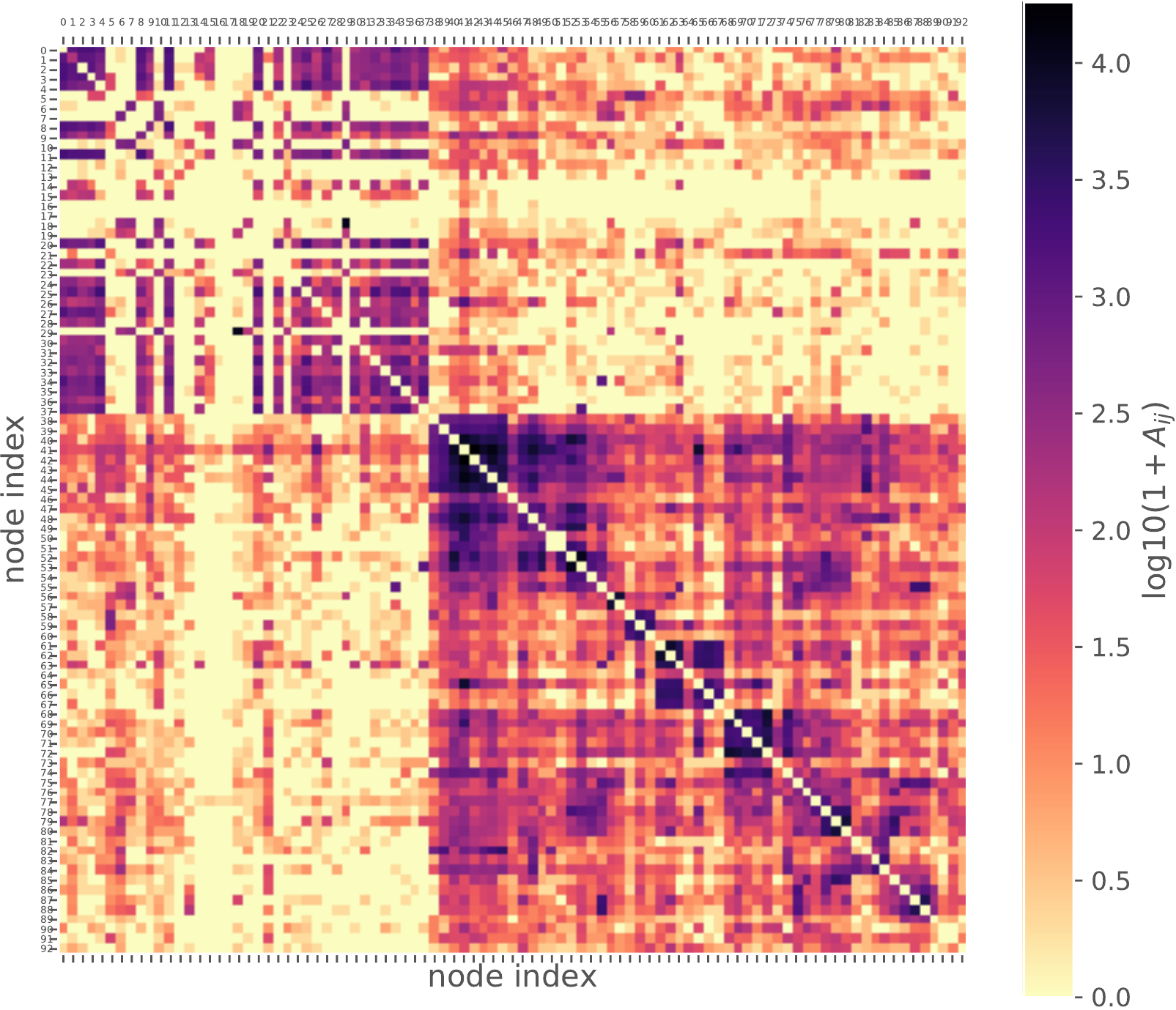}} \hfill
  \subfigure[\label{fig:rm:mask}]{\includegraphics[width=.28\columnwidth]{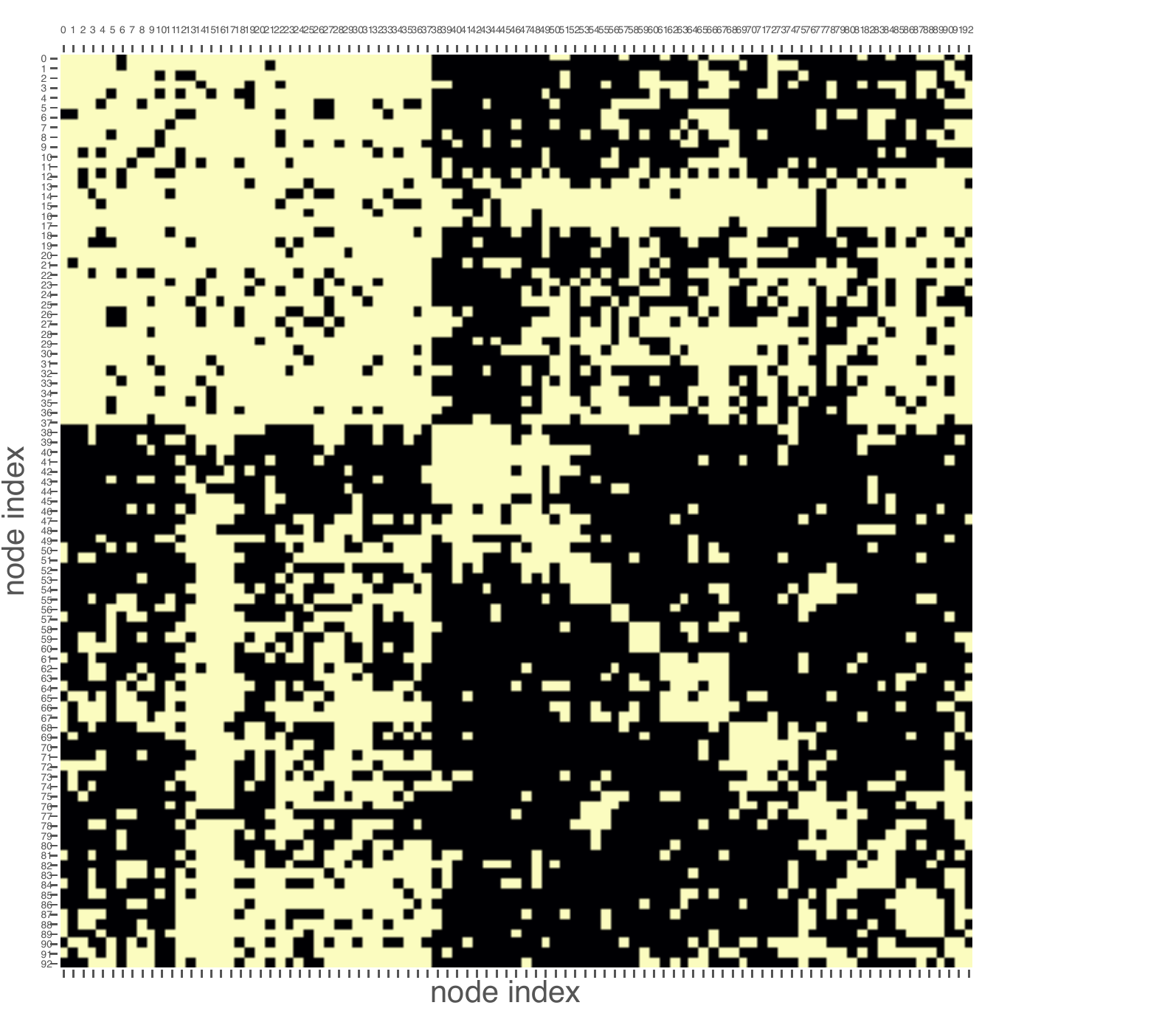}}\hfill
   \subfigure[\label{fig:rm:filtered}]{\includegraphics[width=.28\columnwidth]{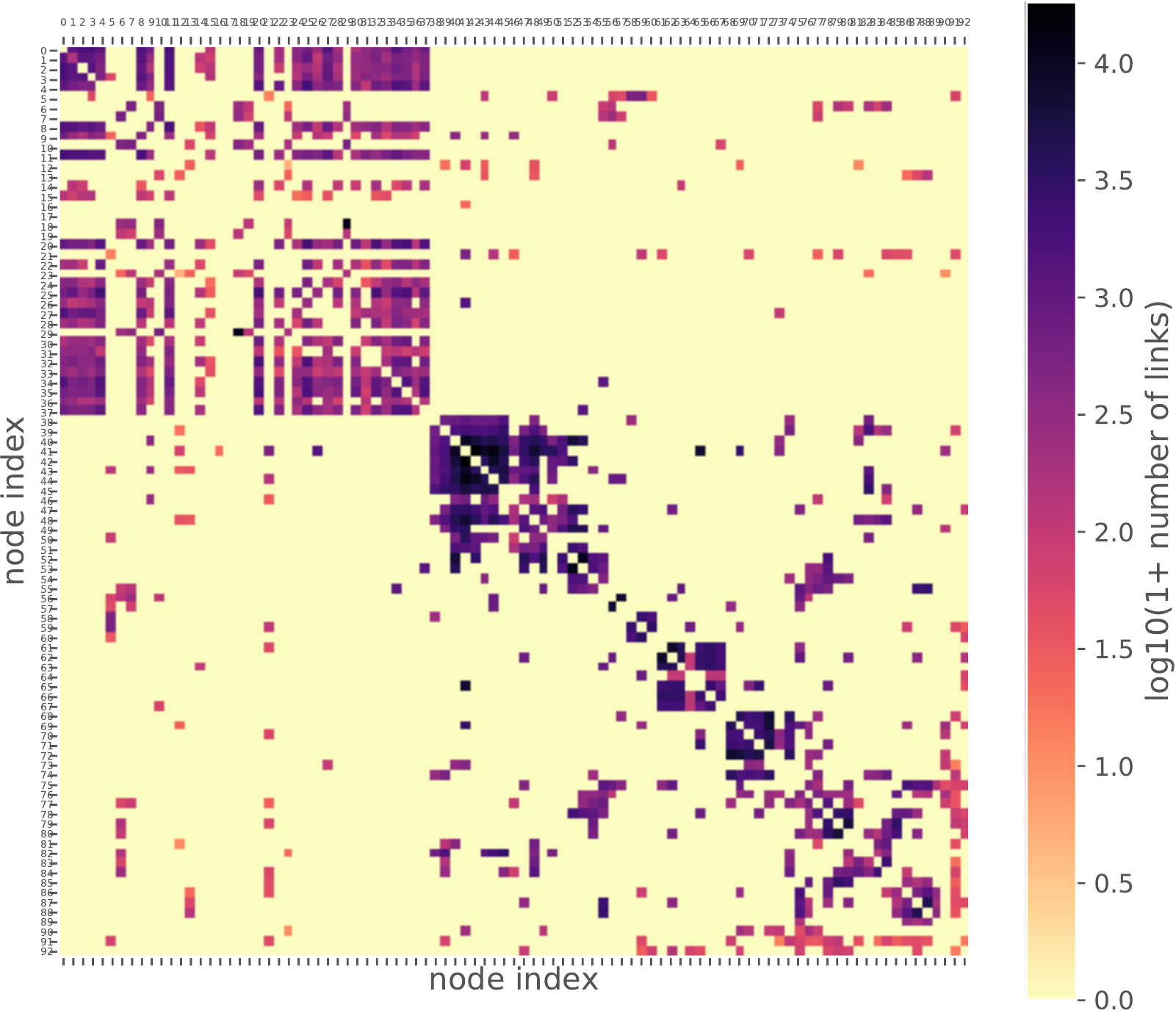}}
   \\
  \subfigure[\label{fig:rm:unfiltered:net}]{\includegraphics[width=.28\columnwidth]{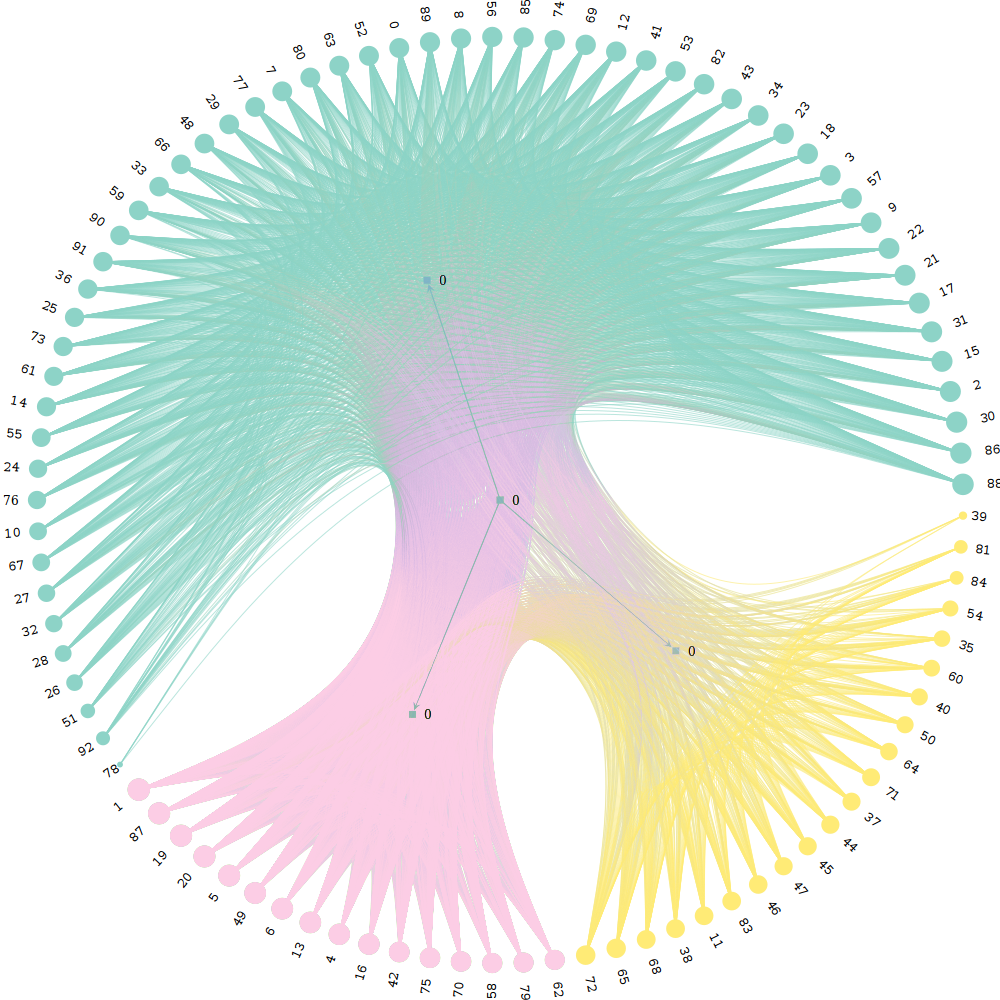}}\hfill\mbox{}
\hfill
  \subfigure[\label{fig:rm:label-match}]{\includegraphics[width=.28\columnwidth,angle=0,origin=c]{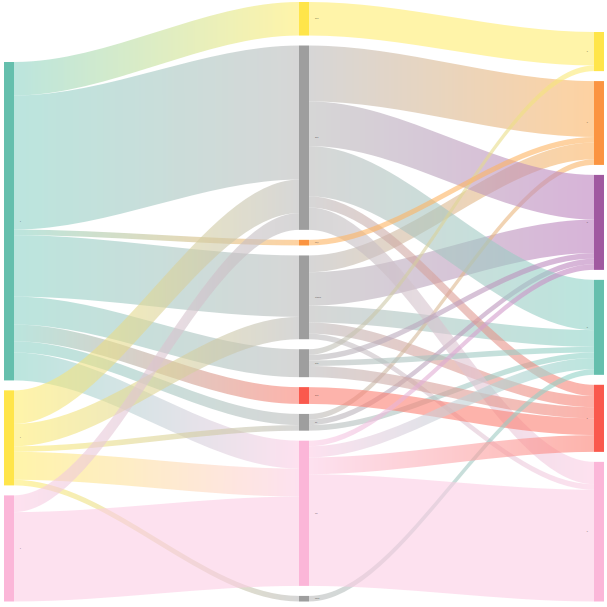}}\hfill\mbox{}
 \hfill
  \subfigure[\label{fig:rm:filtered:net}]{\includegraphics[width=.28\columnwidth]{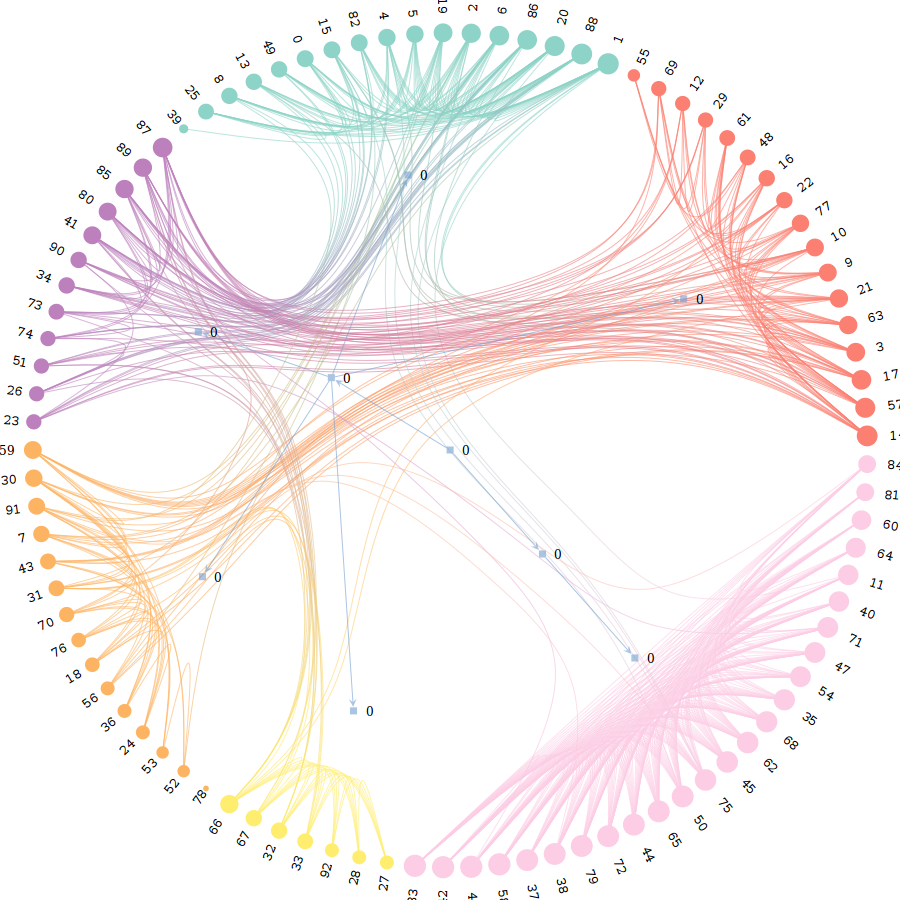}}\hfill\mbox{}
  \caption{  Illustration of our approach in the (RM) data set capturing proximity of students and staff at MIT campus. For the observed weighted adjacency matrix (a) and a given significance threshold, our framework allows to establish a high-pass noise filter matrix (b), which can be used to obtain a filtered adjacency matrix containing only significant links (c).
  A visual comparison of the output of a community detection algorithm on the unfiltered (d) and filtered (f) graphs shows that detected partitions in the filtered one better correspond to ground truth lab affiliations and classes (e).
  \textbf{(a)} Unfiltered weighted adjacency matrix.
  \textbf{(b)} High-pass noise filter matrix.
  \textbf{(c)} Filtered adjacency matrix containing only significant links.
  \textbf{(d)} Unfiltered graph.
  \textbf{(e)} Comparison of ground truth lab affiliations (center column) vs. detected communities in the unfiltered (left column) and filtered (right column) graph.
  \textbf{(f)} Filtered graph.
}
\label{fig:rm}
\end{figure}

To illustrate our approach, Figure~\ref{fig:rm:unfiltered} shows the entries of the (original) adjacency matrix $\mathbf{A}$ for (RM).
The high-pass noise filter resulting from our methodology (using $\alpha=0.01$) is shown in Figure~\ref{fig:rm:mask}, where black entries correspond to pairs of nodes with non-significant links.
The application of this filter to the original matrix yields the noise-filtered matrix shown in Fig.~\ref{fig:rm:filtered}.
While in the full network there are $721,889$ observed multi-edges amounting to $2,952$ distinct links, after filtering there are $626$ $(21.2\%)$ significant links left ($617,069$ multi-edges, $85.5\%$ of the original).
We validate the benefit of filtering the original interactions in (RM) by comparing the output of a standard community detection algorithm -- the degree-corrected block model \cite{Peixoto2014a} -- in (i) the original, unfiltered graph shown in Fig.~\ref{fig:rm:unfiltered:net}, and (ii) the filtered, significant graph shown in Fig~\ref{fig:rm:filtered:net}.
Using known classes of students and affiliations of staff members as ground truth allows us to compare the quality of the community detection.
Figure~\ref{fig:rm:label-match} shows the set overlaps between the ground truth labels (middle column) and detected partitions in the unfiltered (left column) and filtered graph (right column).
Due to the high number of non-significant links in the unfiltered graph, the algorithm only detects three partitions, each spanning multiple labs and classes.
In contrast, applying the algorithm to the filtered graph yields six partitions that better capture the ground truth lab and class structure (cf. Fig.~\ref{fig:rm:label-match}).
As expected, detected partitions do not perfectly correspond to the ground truth, since labs and classes are likely not the only driving force behind observed proximities.

\begin{figure}[htpb]
  \centering
  \subfigure[\label{fig:zkc:unfiltered}]{\includegraphics[height=.4\columnwidth]{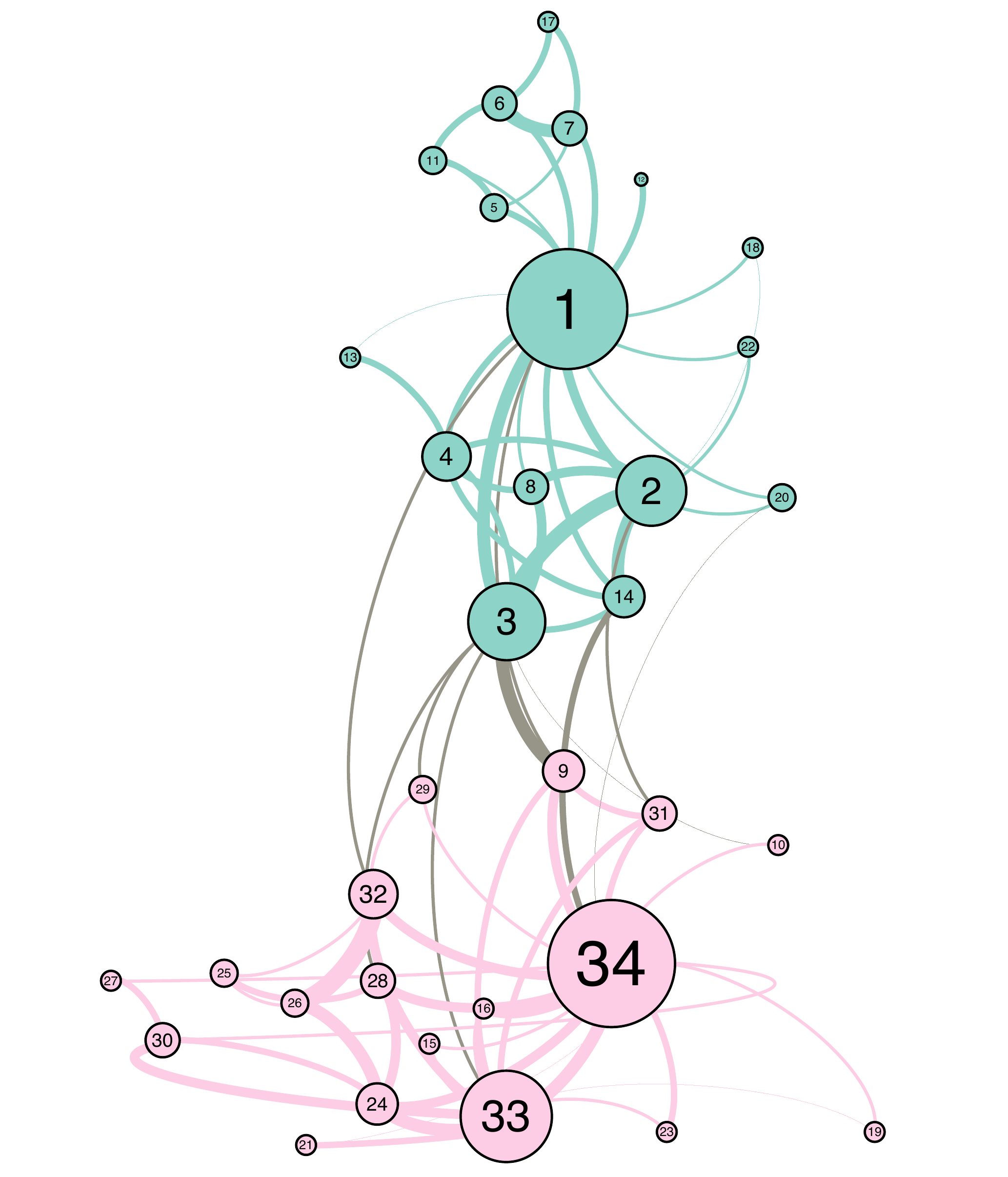}} \hfill
  \subfigure[\label{fig:zkc:filtered}]{\includegraphics[height=.4\columnwidth]{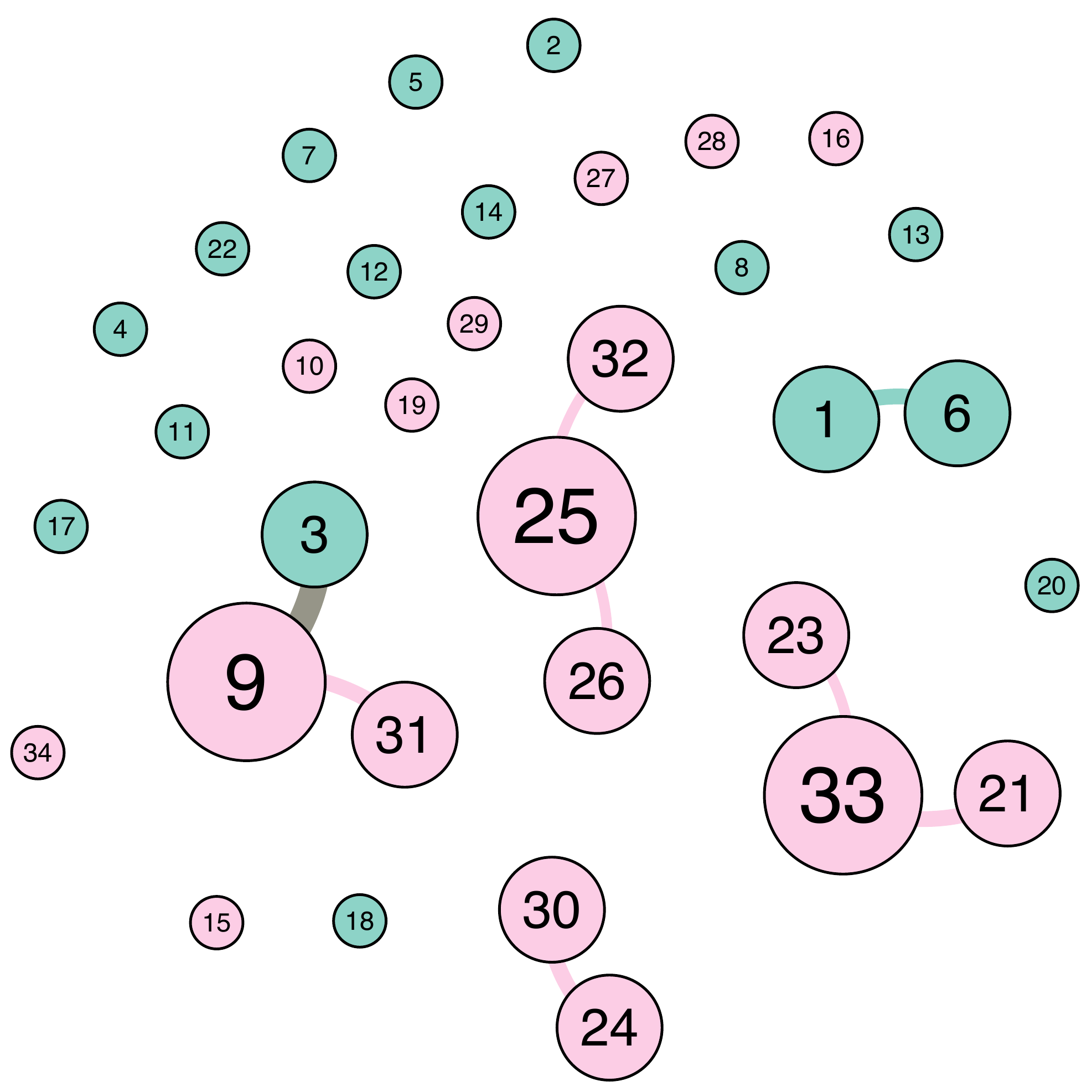}}\hfill\mbox{}
  \caption{Observed (a) and filtered (b) weighted graphs for the (ZKC) data set, capturing encounters between members of a Karate club. The filtered graph shows that most of the observed encounters can be explained by random effects resulting from the club members' separation into two classes.}
\label{fig:zkc}
\end{figure}

A major advantage of gHypEs is that, by specifying a non-uniform matrix $\mathbf{\Omega}$, we can additionally encode known factors that influence the occurrence of interactions between nodes, while still obtaining an analytically tractable ensemble.
In our second illustrative example, we use this to encode the known structure of two separate Karate classes in the (ZKC) data.
These two classes naturally influence the frequency of encounters between actors beyond what would be expected ``at random''.
We incorporate this prior knowledge via a block matrix $\mathbf{\Omega}$ that assigns higher dyadic propensities to pairs of actors in the same class (cf. \cite{Casiraghi2017}).
This approach allows to establish a ``random baseline'' accounting both (i) for combinatorial effects due to heterogeneous node degrees, and (ii) the known group structure in the data.
Using a significance threshold of $\alpha=0.01$, for (ZKC) this yields the striking result that only $8$ out of $78$ observed links are significant ($\sim90\%$ of $231$ observed multi-edges are filtered out, cf. fig.~\ref{fig:zkc}).
In other words, taking into account the partitioning of members in two classes for (ZKC) almost all encounters between club members can simply be explained by random effects.
\Cref{fig:zkc} compares the original weighted network, illustrated in \cref{fig:zkc:unfiltered}, and the filtered network, in \cref{fig:zkc:filtered}.

\vspace{-.2cm}

\section{Conclusion}

In this short paper we introduce gHypEs, a broad class of statistical ensembles of graphs that can be used to infer significant links from noisy data.
Our work makes three important contributions:
First, we provide an analytically tractable statistical model of directed and undirected multi-edge graphs that can be used for inference and learning tasks.
Second, the formulation of our ensemble highlights a -- to the best of our knowledge -- previously unknown relation between random graph theory and Wallenius` non-central hypergeometric distribution.
And finally, different from existing statistical ensembles such as, e.g., the configuration model, our framework can be used to encode prior knowledge on factors that influence the formation of relations.
This flexible approach allows for a tuning of the ``random baseline'', opening perspectives for a statistically principled network inference that accounts for effects that are not purely random.
We thus argue that our work advances the theoretical foundation for the mining of relational data on social systems.
It further highlights that principled model selection and hypothesis testing are crucial prerequisites that should precede the application of network-based data mining and modeling techniques.

\end{document}